\documentclass[preprint,aps,superscriptaddress]{revtex4}

\newcommand{\fig}[1]{Figure \ref{#1}}

\usepackage{amsmath}
\usepackage{epsfig}

\begin{document}
\title{Mechanism of Luminescence Ring Pattern Formation in Quantum Well Structures: Optically-Induced In-Plane Charge Separation}

\author{R. Rapaport\footnote{These authors have equal contributions.}}
\affiliation{Bell Laboratories, Lucent Technologies, 600 Mountain
Avenue, Murray Hill, New Jersey 07974}

\author{Gang Chen$^*$}
\affiliation{Bell Laboratories, Lucent Technologies, 600 Mountain
Avenue, Murray Hill, New Jersey 07974}

\author{D. Snoke$^*$}
\affiliation{Department of Physics and Astronomy, University of
Pittsburgh, 3841 O'Hara St., Pittsburgh, PA 15260}

\author{Steven H. Simon}
\affiliation{Bell Laboratories, Lucent Technologies, 600 Mountain
Avenue, Murray Hill, New Jersey 07974}

\author{Loren Pfeiffer}
\affiliation{Bell Laboratories, Lucent Technologies, 600 Mountain
Avenue, Murray Hill, New Jersey 07974}

\author{Ken West}
\affiliation{Bell Laboratories, Lucent Technologies, 600 Mountain
Avenue, Murray Hill, New Jersey 07974}

\author{Y. Liu}
\affiliation{Department of Physics and Astronomy, University of
Pittsburgh, 3841 O'Hara St., Pittsburgh, PA 15260}

\author{S. Denev}
\affiliation{Department of Physics and Astronomy, University of
Pittsburgh, 3841 O'Hara St., Pittsburgh, PA 15260}



\maketitle

{\bf  About a year ago, two independent
experiments\cite{ButovNature2002, SnokeNature2002}, imaging
indirect exciton luminescence from doped double quantum wells
under applied bias and optical excitation, reported a very
intriguing observation: under certain experimental conditions, the
exciton luminescence exhibits a ring pattern with a dark region in
between the center excitation spot and the luminescent ring that
can extend more than a millimeter from the center spot.  Initial
speculations on the origin of this emission pattern included
supersonic ballistic transport of excitons due to their
dipole-dipole repulsion and Bose superfluidity of excitons. In
this paper we show that the ring effect is also observed in single
quantum well structures, where only direct excitons exist. More
importantly, we find that these experimental results are
quantitatively explained by a novel coupled 2D electron-hole
plasma dynamics, namely, photoinduced in-plane charge separation.
This charge separation explains extremely long luminescence times
that may be more than a microsecond for the ring --- orders of
magnitude longer than the emission lifetime of the excitons in the
center spot. This method of continuously creating excitons may
result in a highly dense exciton gas which is also well
thermalized with the lattice (since the particles can cool over
the very long luminescence time after their hot optical creation),
thus opening up opportunities for a detailed study of quantum
statistics. The in-plane separation of the charges into positive
and negative regions, with a sharp interface between them is an
interesting new example of nonequilibrium dynamics and pattern
formation. \bf}

\begin{figure}[htb]
\begin{center}
\includegraphics[scale=1.5]{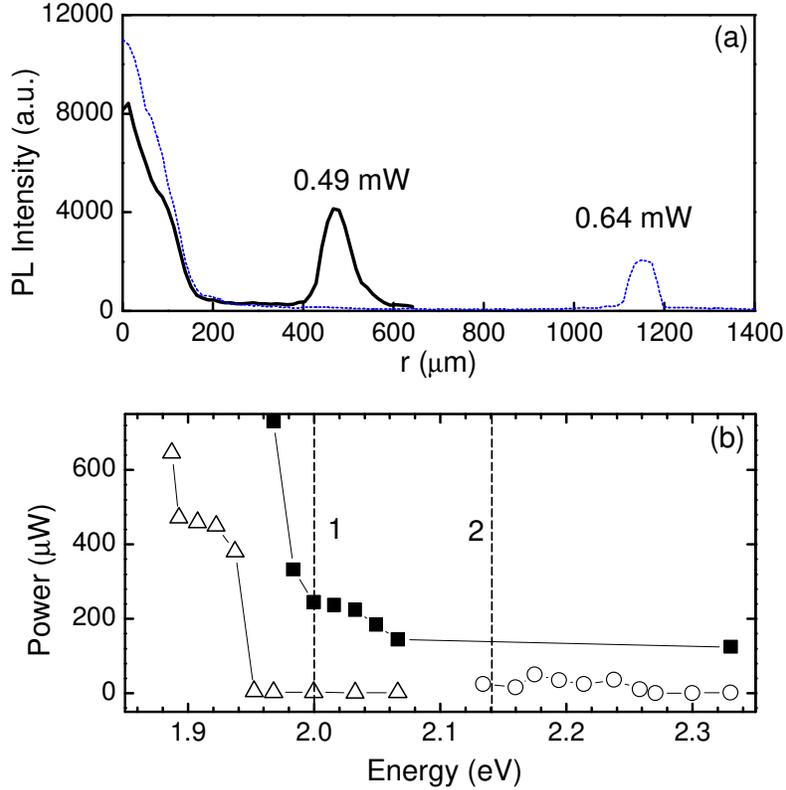}
\caption{(a) Example of the luminescence ring for a double quantum
well structure (Structure A, see below) similar to the one used in
Ref. \cite{SnokeNature2002} for two different laser powers.
 The sample was measured at T=2K, and
excited with a HeNe laser (632 $nm$) with a spot diameter of $\sim
100\mu m$ at $r=0$. (b) Laser power threshold for the formation of
the ring as a function of the excitation energy. This plot shows
that the ring forms easily only for excitation energies above the
bandgap of the barriers. For samples A and B  (solid squares and
open triangles, respectively) the bandgap of the barriers (2.0 eV)
is marked by vertical line 1. For excitation energies below this
threshold, a ring can be formed only by using much larger laser
power.   For sample C (open circles) the bandgap of the barriers
(2.15 eV) is marked with vertical line 2.  For this sample, no
ring was observed for excitation energies below 2.13 eV.  In
samples A and B no ring was observed for any excitation power for
excitation energies below 1.88 eV.
Structure A (solid squares) consists of two 60 \AA\ ${\rm
In}_{0.1}{\rm Ga}_{0.9}{\rm As}$ quantum wells, separated by a 40
\AA\ GaAs barrier, with 300 and 1000 \AA\ ${\rm Al}_{0.32}{\rm
Ga}_{0.68}{\rm As}$ outer barriers separated from the wells by 50
\AA\ GaAs buffer layer. Structure B consists of two 80 \AA\ GaAs
quantum wells, separated by a 40 \AA\ AlAs barrier, with 2000 \AA\
$\rm Al_{0.33}Ga_{0.67}As$ outer barriers, which is the same
design as used in Ref. \cite{ButovNature2002}. Structure C (open
circles) is the same as structure A, but with $\rm
Al_{0.44}Ga_{0.56}As$ outer barriers. Values for bandgaps are
taken from Ref. \cite{Adachi}.} \label{fig0}
\end{center}
\end{figure}

Butov et. al. \cite{ButovNature2002} and Snoke et. al.
\cite{SnokeNature2002} independently studied the emission from a
modulation doped double quantum well (QW) structures under optical
excitation from a tightly focused laser beam. When voltage is
applied in the growth direction, emission is observed from
indirect excitons --- bound states of a single electron in one QW
with a single hole in the other.  Imaging the spatial distribution
of this emission reveals a surprising pattern: in addition to the
expected emission at the excitation spot, an emission ring is seen
at large radial distances, with a dark region in between the
center spot and the ring. The radius of the ring emission depends
on the excitation power. An example of such an emission pattern is
shown in \fig{fig0}a.  This striking effect calls for a consistent
explanation of its underlying physical origin, which is the main
purpose of this paper. Initial speculations based on various
exciton transport mechanisms and macroscopically coherent
excitonic effects are inconsistent with many experimental results
\cite{SnokeSSC2003}. It is obvious that there is transport of
particles from the excitation spot to the remote ring over
macroscopic distances. However, in order to understand this
effect, there are a few key questions one needs to address: (a)
what types of particles or quasi-particles are being transported
through the dark region to the ring location and what is the role
of excitons in the transport and in the emission compared to that
of free carriers? (b) what is the driving force for this unique
transport?

One of the first key clues for solving this problem is our
observation that identical ring patterns are also seen in single
quantum well structures (see below), as opposed to the double QW
structures studied in the initial reports \cite{ButovNature2002,
SnokeNature2002}. In our structure, there are only direct excitons
(bound states of a single electron and a single hole in the {\it
same} QW), which implies that the effect does not depend on the
specific characteristics of indirect excitons (such as long
lifetimes) as was suggested previously. Another important clue can
be found in the observed excitation energy threshold of the ring
formation as shown in \fig{fig0}b: the ring easily forms only with
photoexcitation energies which are above the AlGaAs barrier
bandgap energy. This implies that the photogenerated hot carriers
are a crucial ingredients in the ring formation and that the
effect depends on the structure as a whole and not only on the
local QW region. Since it is known that in modulation doped
structures, optical excitation above the barriers can lead to
depletion of electrons \cite{KukushkinPRB1989}, it is interesting
to ask whether the mechanism responsible for the ring formation
has to do with such optical depletion.

It is then sensible to obtain further insight into the above
mentioned questions by performing careful spectroscopic studies.
The lineshapes, linewidths, and energies of the emission from the
center spot and the ring are very good local probes of the carrier
density distributions in the QW plane.  In particular, excitonic
emission lines characteristic of low densities should be
significantly narrower, more symmetric, and blue shifted compared
to emission resulting from recombination of free electron-hole
plasmas of high densities \cite{Ronen}.

\begin{figure}[htb]
\begin{center}
\includegraphics[scale=0.7]{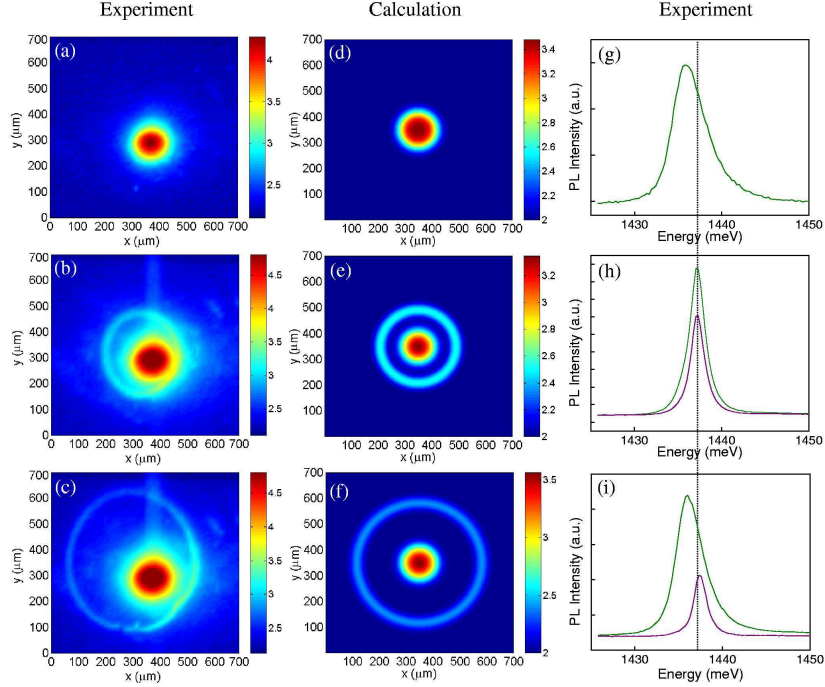}
\caption{Photoluminescence images for our single quantum well
sample taken at three different excitation powers, 50, 265 and 296
$\mu W$, are shown in the left panel (a)-(c). A ring emission
pattern forms when the power is larger than a threshold. The
middle panel (d)-(f) shows the results of the model calculations
based on a photo-induced, in-plane charge separation (see text).
The parameters used for the calculations are discussed in details
in the methods section.  The asymmetry of the ring pattern with
respect to the center spot is probably due to a gradient of the
barrier width, which is not taken into account in our model. The
right panel (g)-(i) presents the experimentally observed center
spot (olive) and ring (purple) emission spectra. In (g) and (i) we
see a broadening and a red-shift of the center spot which
indicates a high density of carriers compared to the center spot
in (h) and to the ring in (h) and (i).  The sample was measured at
T=12K, and excited with a HeNe laser (632 $nm$) with a spot
diameter of $\sim 60\mu m$. Our sample consists of a  single
60\AA\ In$_{0.13}$Ga$_{0.87}$As quantum well surrounded by
GaAs/Al$_{0.32}$Ga$_{0.68}$As 50\AA/1000\AA\ barriers.
 A 1000 \AA\ layer of Si doped GaAs is located 2000 \AA\ from the QW on the $n^+$ substrate
 side and another similar layer is located 1000\AA\ from the QW on the top contact side.
See \fig{fig3}a for a schematic diagram of the QW structure. Gold
films are deposited on both sides of the sample to form contacts
for applying bias. A 3mm hole is opened on the top gold film for
the optical measurements. } \label{fig1}
\end{center}
\end{figure}
\begin{figure}[htb]
\begin{center}
\includegraphics[scale=1.4]{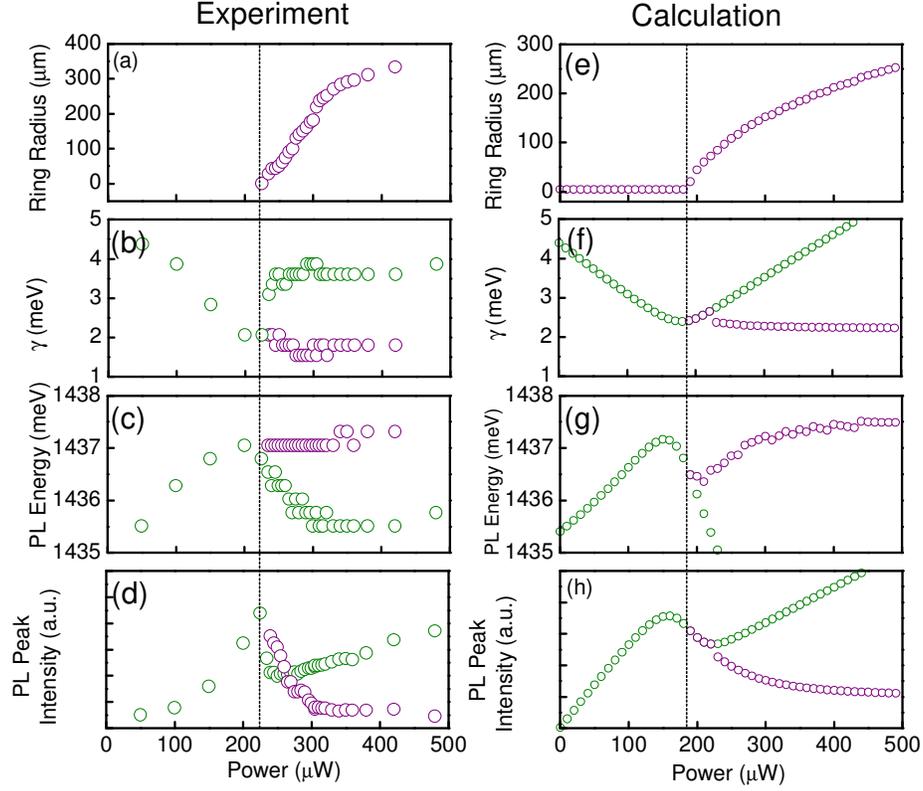}
\caption{Comparison between the experimental and model calculated
excitation power dependences: (a) experimental and (e) calculated
ring radius. The ring radius increases sublinearly after the
excitation power reaches beyond a certain threshold. (b) the
linewidth and (c) the energy of the center spot luminescence
(olive circles) and the ring luminescence (purple circles). (f)
Model calculated linewidth of the luminescence of the center spot
and ring assuming degenerate 2D electron-hole plasma. (g) Model
calculated energies of the center spot and ring luminescence based
on bandgap renormalization. (d) Experimental and (h) calculated PL
peak intensity at the center spot and the ring. Vertical dashed
lines mark the onset of the ring formation.
 Note: The ring radius is always measured in the -y
direction.}\label{fig2}
\end{center}
\end{figure}

\fig{fig1}a-c show the ring pattern from our single well sample at
various different excitation powers. As shown in more detail in
\fig{fig2}a, the ring radius grows (sublinearly) with increasing
excitation power above a certain critical threshold for ring
creation. \fig{fig1}g-i show the center spot and ring PL spectra
corresponding to \fig{fig1}a-c. The detailed behavior of the
spectral linewidth and energy as a function of excitation power is
shown in \fig{fig2}b-c for both the center spot and the ring. From
\fig{fig2}b-c, we see the following trend: At low excitation
power, there is no ring and the emission of the center spot is
broad, asymmetric, and red shifted (see also \fig{fig1}g), which
would be characteristic of emission from a region with a high
density of dissociated charge carriers (an electron-hole plasma).
As the power is increased towards the critical power for ring
creation, the linewidth of the central spot narrows by a factor of
two, becomes more symmetric (not shown), and less red shifted
compared to the central spot at lower density. Since the line is
symmetric and narrow, we believe this to be the natural emission
of a gas of excitons. At excitation powers just above the critical
power for ring creation, the spectra of both the ring and the
central spot remain narrow and symmetric, and remain roughly at
the same energy (see also \fig{fig1}h). As the power is further
increased and the radius of the ring grows, the spectrum of the
central spot red shifts, broadens, and becomes asymmetric (see
also fig \ref{fig1}i), indicative of a gas of dissociated carriers
again. In contrast, the spectrum of the ring remains nearly
constant: unshifted, symmetric, and narrow, indicative of a gas of
excitons.  Finally, \fig{fig2}d plots the peak PL intensities as a
function of power. The center spot emission increases with power,
except for a sudden drop around the threshold for ring creation.
The ring intensity, on the other hand, drops with increasing power
(and corresponding increasing radius).

\begin{figure}[htb]
\begin{center}
\includegraphics[scale=0.7]{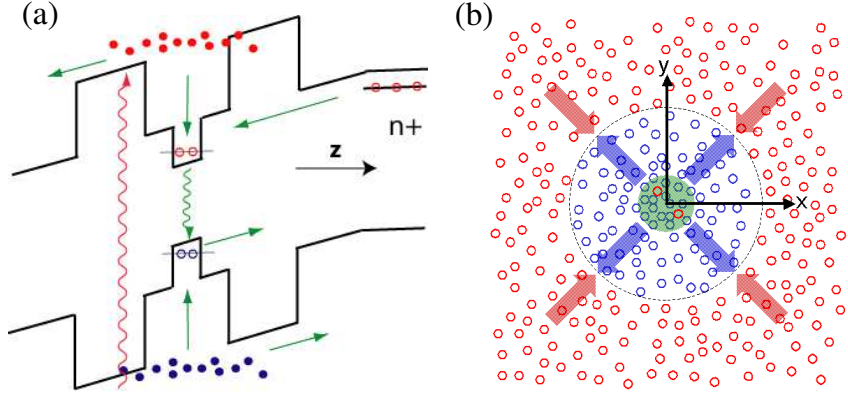}
\caption{(a) schematically describes the energy profile of the
quantum well structure in the growth direction. Without light, the
quantum well contains a 2D electron gas due to modulation doping
with a density, and leakage current that depends on the bias. In
our model, hot electrons (red solid circles) and holes (blue solid
circles) generated by light with an energy above the barriers can
either drift to the contacts or cool to become trapped in the
quantum well (cold carriers - open circles). The hot holes have a
higher probability of cooling down than hot electrons (see text).
The excess of cold holes can recombine with the cold electrons
leading to a depletion of cold electrons  and accumulation of cold
holes at the excitation spot.  Cold electrons and cold holes can
also leak to/from the contacts in this model.
 (b) Excess cold holes (open blue
circles) built up near the excitation spot (shaded circular area)
diffuse outwards while the cold electrons (open red circles),
present in the absence of photoexcitation, far from the the
excitation spot, diffuses inward.  A  sharp boundary (dotted
circle) is formed between the hole puddle and the sea of electrons
surrounding it. The recombination of electrons and holes at the
boundary gives rise to the ring pattern.  There is also a low
density of photoexcited electrons in the center spot which quickly
recombine with holes.}\label{fig3}
\end{center}
\end{figure}

We now try to construct a consistent physical picture to account
for all of our experimental results. \fig{fig3}a shows a schematic
description of the relevant processes associated with
above-the-barrier photoexcitation, under an applied bias in the
growth direction. Without photoexcitation, a 2D electron gas is
present in the QW due to the modulation doping of the sample (this
is likely to be true even for the samples of Ref. \cite{
ButovNature2002} when under bias).   Applying bias will change the
electron density in the QW and generate a leakage current. When
hot electrons and holes are photogenerated by the laser, as shown
by the red and blue dots in \fig{fig3}a, they can either drift
under the applied bias to the contacts or cool down through
phonon-carrier or carrier-carrier scattering, and consequently get
trapped in the QW (red and blue open circles). However, the
trapping rate of holes is expected to be larger than that of the
electrons. This can be attributed to several factors: First, the
hole has a much smaller drift velocity (mobility) than the
electron, by a factor of 10-20 \cite{adachibook}, so it takes
longer for the hole to drift away from the QW, and thus it has a
higher probability of scattering into the QW. Secondly, the phonon
scattering time of the holes is much shorter than that of the
electrons \cite{shahbook}. This results in the hot holes cooling
faster (and drifting slower). As a consequence, there is an excess
of photogenerated holes that end up trapped in the QW. These holes
will recombine with the electron gas in the QW, thus depleting it.
The degree of depletion depends on the photogeneration rate
compared to the leakage rate from the contacts which can replenish
the electron density. With high enough excitation power, the
electron gas can be almost completely depleted, leaving a
population of trapped cold holes instead. Since the illumination
area is small, the optically induced depletion of electrons and
the accumulation of the holes in the QW plane is local. As a
result, close to the center spot, there will be a puddle of holes,
surrounded by a sea of electrons. Due to diffusion, the holes tend
to spread outwards, while the electrons will flow inwards. This
process is schematically shown in \fig{fig3}b. In steady state,
the holes and electrons are spatially separated in the plane of
the QW, \textit{perpendicular} to the applied electric field,
forming a sharp (nominally circular) boundary which extends to a
radius larger than that of the excitation spot. Since electrons
and holes can only recombine where they meet, a sharp luminescence
ring will form at the boundary between the two regions of opposite
charges.

With this physical picture, the dynamics of the carriers in the QW
plane can be described using the following coupled rate equations:
\begin {eqnarray}
\frac{\partial n_{hot}}{\partial t}&=&D_{hot}^e\nabla^2n_{hot}-
\frac{n_{hot}}{\tau_{cool}^e}-\frac{n_{hot}}{\tau_{drift}^e}+Af(r)
\label{nhot}\\
\frac{\partial p_{hot}}{\partial t}&=&D_{hot}^h\nabla^2p_{hot}
-\frac{p_{hot}}{\tau_{cool}^h}-\frac{p_{hot}}{\tau_{drift}^h}+Af(r)\label{phot}\\
\frac{\partial n_{cold}}{\partial
t}&=&D_{cold}^e\nabla^2n_{cold}+\frac{n_{hot}}{\tau_{cool}^e}-\frac{n_{cold}-n_{eq}}{\tau_{leak}^e}
-\xi \,
n_{cold} \,\, p_{cold} \label{ncold}\\
\frac{\partial p_{cold}}{\partial
t}&=&D_{cold}^h\nabla^2p_{cold}+\frac{p_{hot}}{\tau_{cool}^h}-\frac{p_{cold}}{\tau_{leak}^h}
-\xi \, n_{cold}\,\, p_{cold} \label{pcold}
\end {eqnarray}

The first two equations describe the dynamics of the density
distributions of hot electrons $n_{hot}(\vec r)$ and hot holes
$p_{hot}(\vec r)$ in the plane.   Here, the first term on the
right hand side of these equations allows the hot electrons and
holes to diffuse with diffusion constants $D^e_{hot}$ and
$D^h_{hot}$ (the $\nabla^2$ are derivatives in the in-plane
directions only).   The next two terms on the right hand side
allow the hot electrons and hot holes to cool with characteristic
times $\tau_{cool}^e$ and $\tau_{cool}^h$ respectively or to drift
to the contacts with times $\tau_{drift}^e$ and $\tau_{drift}^h$.
Finally, we have added the source term $A f(r)$ which creates hot
electrons and holes from photons.  Here, $f(r)$ is the normalized
excitation beam profile and $A$ is the total absorbed photon flux
(each absorbed photon generates one electron and one hole).

Similarly, the third and fourth equations describe the dynamics of
the density distributions of cold electrons $n_{cold}(\vec r)$ and
cold holes $p_{cold}(\vec r)$ in the plane.  Again, the first term
on the right hand side allows the  cold electrons and holes to
diffuse with diffusion constants $D^e_{cold}$ and $D^h_{cold}$.
The second term on the right hand side generates cold carriers
from any hot carriers that are cooled down.  The third term on the
right hand side allows the cold electrons and cold holes to leak
to or from the contacts with rates $\tau_{leak}^e$ and
$\tau_{leak}^h$ respectively, in a way that tries to bring the
densities back to the (equilibrium) densities of the dark state
($n_{eq}$ for electrons, and zero for holes). Finally, the last
term on the right hand side represents the recombination of cold
electrons and cold holes, where $\xi$ is the electron-hole capture
(or collision time) coefficient.  Note that we have written this
term as being proportional to the product $n_{cold} \, p_{cold}$
which assumes that the rate limiting step in recombination is the
rate for the electron and the hole to find each other and form a
pair, {\it i.e.}, we have assumed that once an electron-hole pair
is formed, it will recombine immediately.  Note also that the
intensity of emission is given by the number of electrons and
holes that recombine and hence the spatial profile of emission is
given by the spatial dependence of the product $n_{cold} \,
p_{cold}$.

Although there are many parameters in this model, all of them are
known or can at least be estimated. A detailed discussion of all
these parameters is given in the methods section below. Further
simplification of these equations can be made by neglecting the
diffusion of hot carriers in the plane. This is a reasonable
assumption because in-plane diffusion is typically much slower
than the drifting and cooling processes. Under these assumptions,
the equations for the hot carriers can be solved exactly. This
reduces the problem to a set of only two coupled rate equations
for the cold carriers (see methods section below).

\begin{figure}[htb]
\begin{center}
\includegraphics[scale=1.5]{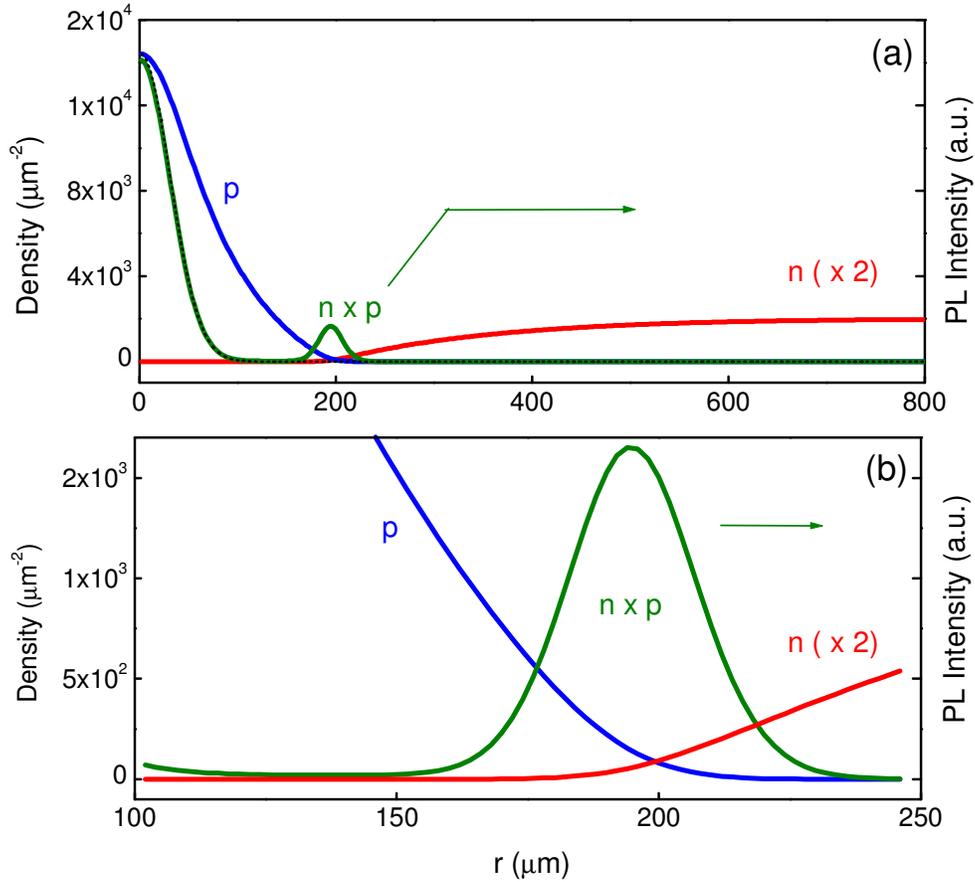}
\caption{(a) Model calculation for a continuous-wave excitation at
a power of $350 \mu W$. The black dotted line shows the incident
photon intensity radial profile. The red and blue lines show the
cold electron and cold hole density radial profiles, respectively.
The electron depletion and hole accumulation regions are clearly
seen to be larger than the excitation spot.  The photoluminescence
intensity profile, which is proportional to the product of the
cold electron and cold hole densities, is plotted in olive. The
ring emission pattern forms at the boundary between the puddle of
cold holes and the sea of electrons. (b) is a expanded view of (a)
around the ring position.} \label{fig4}
\end{center}
\end{figure}

The steady state solutions of the model are presented in
\fig{fig4} for the set of parameters discussed in the methods
section below at an excitation power level where a ring is formed.
The laser intensity profile is shown in by the black dotted line.
The in-plane cold electron and hole density distribution profiles
are shown by the red and blue lines, respectively.  Consistent
with our qualitative picture, the model predicts that both the
electron depletion region and the puddle of holes are larger than
the excitation spot due to the diffusion processes. The olive line
shows the emission profile, which is just the product of the cold
electron and cold hole densities.  The emission ring appears at
the sharp boundary between the electron sea and the hole puddle.
The calculated emission patterns corresponding to the experimental
conditions of \fig{fig1}a-c are plotted in \fig{fig1}d-f. There is
a good agreement between the experiment and calculation except for
the experimental asymmetry of the ring with respect to the center
spot. We attribute the asymmetry of the ring to a gradient of
barrier width, which for simplicity is not taken into account in
our model.   As the barrier width changes there will be
corresponding changes in $n_{eq}$ causing the ring radius to
locally vary.

In order to check the validity and consistency of our model, we
compare its predictions to our spectroscopic experimental findings
in \fig{fig2}. First, as the excitation power is increased above a
threshold, the electron depletion region and the area of the hole
puddle increase. This pushes the ring out and away from the center
spot. This behavior is well reproduced by the model, as seen in
\fig{fig2}e. The second crucial test of our physical picture is
its ability to explain the puzzling dependence on excitation power
of the linewidth and energy of the PL of the center spot. It is
well known that the emission linewidth of a 2D electron-hole gas
is a measure of the holes and electron Fermi energies which are
determined by their corresponding densities ($n_{cold}$ and
$p_{cold}$). \fig{fig2}f shows the calculated linewidth ($\gamma$)
assuming a quasi-degenerate 2D electron-hole gas, and is given by
$\gamma=E_{f}^e+E_{f}^h=\pi\hbar^2(n_e/m_e+n_h/m_h)$. Here,
$E_f^e$ and $E_f^h$ are the electron and hole Fermi energies, and
$m_e$ and $m_h$ are the electron and hole effective masses,
respectively. The energy shift due to bandgap renormalization also
depends on $n_{cold}$ and $p_{cold}$: the larger the total carrier
density, the larger the emission red-shift due to carrier-carrier
correlations. Following Ref. \cite{TranklePRB1987,SchmittSSC}, the
bandgap renormalization is approximated by $E_{BGR}=E_0-\eta
(n_{cold}+p_{cold})^{1/3}$, with $\eta$ a fit parameter, and is
plotted in \fig{fig2}g. Here, $E_0$ is the QW transition energy
(at zero density) and $\eta$ is a fitting parameter. Both the
calculated linewidth and energy shift agree well with the
experimental results.  Finally, \fig{fig2}g plots the calculated
peak intensity of the center spot and ring PL as a function of
power. As with the previous dependencies, it agrees well with the
experimental behavior shown in \fig{fig2}d.

In addition to the above described quantitative agreement of the
experiment and the model, it is also very easy to understand how
the qualitative trends come about.  At low excitation power, we do
not generate enough cold holes to substantially deplete the
existing cold electron density (estimated to be $10^{11}cm^{-2}$
from PL linewidth). Thus, we should see PL corresponding to
recombination in the presence of a high density of free charge
carriers which should be broad, red-shifted, and asymmetric, as we
indeed observe (see \fig{fig1}g). Increasing excitation power, the
depletion of electrons becomes more significant, leading to a
reduction of the excess electron density with only a small
increase of holes. This lower density leads to a decrease in
linewidth and a smaller red-shift of the center spot. As the power
is further increased, the center spot is completely depleted of
electrons and the ring starts to appear. At this point, the total
carrier density at the center is around $2\times 10^{10} cm^{-2}$,
and correspondingly the spectral lines are narrower and less
redshifted (see \fig{fig1}h). Increasing the power still further
will start to build up a large hole density in the center spot.
The increase in charge carriers in the center spot then reverses
the previous trend making the central spot once again broad,
asymmetric, and redshifted (see \fig{fig1}i).  In contrast, the
total carrier density at the ring (estimated to be
$\sim2\times10^{10}cm^{-2}$) remains nearly independent of the
excitation power, and therefore no change in its linewidth and
energy position is observed. At this low density, the emission
from the ring appears excitonic.

The ability of our model to describe the various experimental
results strongly supports the validity of the physical picture
presented above. The agreement of the experiment and theory is
impressive --- particularly given the very crude nature of the
theoretical model.  Furthermore, the qualitative trends of the
model are quite independent of the precise values of the input
parameters.

We are now in a good position to answer the questions posed at the
beginning of this paper. According to our model, the particles
that are being transported are free carriers and no additional
excitonic transport mechanism is necessary in order to explain our
experimental data. Nonetheless, it seems that the ring emission is
due to excitonic recombination (as is the center spot for
excitation powers close to the threshold power for ring creation).
The actual thermodynamic state of the exciton gas in the ring
remains a very interesting open question. It was hypothesized in
Refs. \cite{ButovNature2002,SnokeNature2002} that the exciton gas
is statistically degenerate. This is not excluded by our model
(see also the discussion below). In fact, our model predicts a
carrier density of $\sim 2\times 10^{10}cm^{-2}$, which is of the
order of the Kosterlitz-Thouless 2D transition density to a
superfluid state at T=2K. (The data of \fig{fig1} and \ref{fig2},
however, are performed at higher temperatures).   As for the
driving force, diffusion of free carriers seems to sufficiently
account for the observed effects. At first glance, it might seem
that the carrier-carrier Coulomb interaction should play an
important role. However, since the distance of the GaAs n${}^+$
conducting layers from the 2D gas layer is very small, any long
range Coulomb interactions will be screened out, although plasmons
can still propagate, which may be important for understanding the
short time behavior reported in Ref. \cite{SnokeSSC2003}.
Furthermore, the Coulomb term may be important in understanding
the pattern formation which occurs at very low temperatures,
reported in Ref. \cite{ButovNature2002}. In particular, below
T=1.8K, the luminescence ring breaks into a periodic bead-like
structures, which could possibly be understood as pattern
formation arising from competition between repulsion of like
charges and attraction of opposite charges.

While our model explains the experiments described here and in
Ref. \cite{SnokeNature2002,SnokeSSC2003} quite well, there are few
points left unresolved. First, the observed saturation of the
center spot linewidth and energy shift at very high excitation
powers cannot be reproduced by our simple model. At such high
powers, other effects not included in our model, such as heating
and charge buildup, can become significant. Furthermore, the model
can only partially account for the results in Ref.
\cite{ButovNature2002}. For example, neither our experiments nor
our model calculations show any clear evidence of the inner ring
seen by Butov et. al. which may be related to the inability of
excitons to recombine until they have cooled to energies within
the light-cone\cite{ButovNature2002}.

Another point worth mentioning is the dependence of the ring PL
spatial sharpness with temperature. The model predicts that the
ring width increases with increasing particle capture time, which
also agrees with the observation that the ring broadens as the
temperature is raised.  (In the methods section we describe how
the particle capture time is estimated).  However, other
parameters, such as the time constants, may also be temperature
dependent making quantitative comparisons difficult at this stage.

\begin{figure}[htb]
\begin{center}
\includegraphics[scale=0.5]{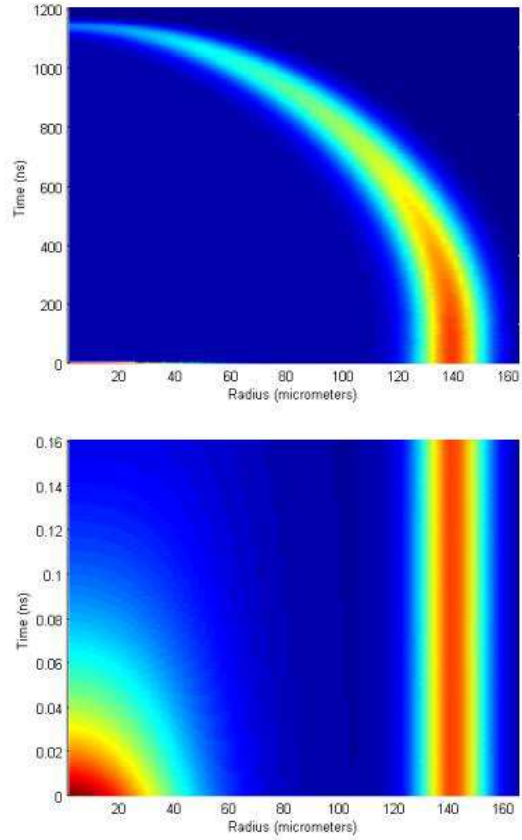}
\caption{(a) Calculated emission pattern cross section as a
function of time after the excitation power is turned off. The
ring emission intensity as well as its radius decay on a time
scale of a microsecond. (b) An expanded view of (a) for very short
times immediately after the power is turned off. The center spot
decays in a typical recombination time of $\sim 50ps$, whereas the
ring emission is very robust and persists for more than a
microsecond after the excitation power is turned off. This time
dependence of our model is in agreement with the experimental time
behaviors reported in Ref. \cite{SnokeNature2002}}\label{fig5}
\end{center}
\end{figure}

A very intriguing prediction of the model is that the macroscopic
charge separation and therefore the ring emission persists for an
extremely long time compared to the center spot lifetime after the
laser excitation is turned off. This behavior is shown in
\fig{fig5}, where the ring lifetime is longer than $1\mu s$. In
comparison, the center spot decays after less than $1ns$. While
the center spot lifetime is determined by the free carrier
recombination time, the ring lifetime is determined by a slow
carrier diffusion, which is driven by the carrier density
gradients in the plane. This long lifetime of the ring is
consistent with time-resolved measurements presented in Ref.
\cite{SnokeNature2002}, which reported a ring PL lifetime longer
than $260ns$.   This long lifetime suggests that excitons being
formed in the ring should be well thermalized because the carriers
should have had plenty of time to cool over the long time from
their hot optical creation.  We suggest that this should be a very
good method of forming cold excitons, opening up opportunities for
studying their quantum statistics at low temperatures.

{\bf{Methods}}

With the above mentioned approximation that diffusion of hot
carriers can be neglected, in steady state the reduced set of
couple equations are simply Eq. 3 and Eq. 4 where
$n_{hot}/\tau^e_{cool} =C_e A f(r)$ and $p_{hot}/\tau^h_{cool}
=C_h A f(r)$ with
$C_{e(h)}=1/(1+\tau_{cool}^{e(h)}/\tau_{drift}^{e(h)})$.   The
ratio between the hot carrier cooling time and drifting time,
$C_e(h)$, can be estimated by comparing the experimentally
measured light induced electric current to the number of
photogenerated hot carriers, which is determined by the number of
photons absorbed in the QW structure.  We use $C_e\cong 0.2$ and
$C_h\cong 1$. We use a gaussian beam profile ($f(r)$) to describe
our excitation spot with a beam diameter of $60\mu m$ FWHM. The
typical photon flux absorbed is $2\times 10^5 ns^{-1}\mu m^{-2}$.
$\xi$, the capture coefficient, is estimated by a simple,
classical free electron-hole Coulomb capture model. In this model,
a charged carrier is assumed to be captured by an opposite charge
if its thermal kinetic energy is smaller than the coulomb
attraction.  This model results in a capture cross section of
$\sigma=\frac{e^2}{6\pi \varepsilon kT}$. $\xi$ is then given by
$\sigma
v_{th}/d=\frac{e^4}{36\pi\varepsilon^2d}\sqrt{\frac{3}{\mu^*k^3T^3}}$
where $v_{th}$ is the thermal velocity of the carrier, $\mu^*$ is
the reduced electron-hole effective mass, and $d$ is the thickness
of the QW. The electron-hole pair capture time is typically much
longer than the recombination time at all the experimental carrier
densities. For example, at a carrier density of $10^{11}$, the
capture time is 1ns. Therefore, the assumption that $\xi$ is
limited by the capture time is well justified. Although this
calculation is quite crude, the calculated capture time is also
consistent with the luminescence rise time of the center spot
which was measured to be $\sim 1ns$ \cite{SnokeNature2002}. The
diffusion coefficient for the electrons is calculated from the
measured electron mobility of the structure to be $D_e=20\mu
m^2/ns$ and that of the holes is then inferred to be $D_h=D_e
m_e/m_h = 5\mu m^2/ns$. The leakage time $\tau_{leak}$ is taken to
be on the order of $10\mu s$, and $n_{eq} = 10^{11}/cm^2$.

{\bf{Acknowledgement}}

This work has been partially supported by the National Science
Foundation and the Department of Energy.  We thank Phil Platzman,
Girsh Blumberg, Nikolai Zhitenev, Rafi de Picciotto, Chandra Varma
and Peter Littlewood for helpful discussions.

{\bf{Competing interests statement}}

The authors declare that they have no competing financial
interests.

Correspondence and requests for materials should be addressed to
S. H. S. (email: shsimon@lucent.com)

\bibliography{bibliography}

\begin{thebibliography}{10}
\expandafter\ifx\csname natexlab\endcsname\relax\def\natexlab#1{#1}\fi
\expandafter\ifx\csname url\endcsname\relax
  \def\url#1{\texttt{#1}}\fi
\expandafter\ifx\csname urlprefix\endcsname\relax\def\urlprefix{URL }\fi

\bibitem[{Butov \emph{et~al.}(2002)Butov, Gossard \& Chemla}]{ButovNature2002}
Butov, L.~V., Gossard, A.~C. \& Chemla, D.~S.
\newblock Macroscopically ordered state in an exciton system.
\newblock \emph{Nature} \textbf{418}, 751--754 (2002).

\bibitem[{Snoke \emph{et~al.}(2002)Snoke, Denev, Liu, Pfeiffer \&
  West}]{SnokeNature2002}
Snoke, D., Denev, S., Liu, Y., Pfeiffer, L. \& West, K.
\newblock Long-range transport in excitonic dark states in coupled quantum
  wells.
\newblock \emph{Nature} \textbf{418}, 754--757 (2002).

\bibitem[{Snoke \emph{et~al.}(2003)Snoke, Denev, Liu, Pfeiffer \&
  West}]{SnokeSSC2003}
Snoke, D., Denev, S., Liu, Y., Pfeiffer, L. \& West, K.
\newblock Luminescence rings in quantum well structures.
\newblock \emph{Solid State Comm.} \textbf{127}, 187--196 (2003).

\bibitem[{Kukushkin \emph{et~al.}(1989)Kukushkin, von Klitzing, Ploog,
  Kirpichev \& Shepel}]{KukushkinPRB1989}
Kukushkin, I.~V., von Klitzing, K., Ploog, K., Kirpichev, V.~E. \& Shepel,
  B.~N.
\newblock Reduction of the electron density in GaAs-Al$_x$Ga$_{1-x}$As single
  heterojunctions by continuous photoexcitation.
\newblock \emph{Phys. Rev. B} \textbf{40}, 4179--4182 (1989).

\bibitem[{Ramon \emph{et~al.}(2002)}]{Ronen}
Ramon, G. \emph{et~al.}
\newblock Scattering of polaritons by a two-dimensional electron gas in a
  semiconductor microcavity.
\newblock \emph{Phys. Rev. B} \textbf{65}, 085232 (2002).

\bibitem[{Adachi(1993)}]{adachibook}
Adachi, S. (ed.).
\newblock \emph{Properties of Aluminum Galium Arsenide} (INSPEC, the Institute
  of Electrical Engineers, London UK, 1993).

\bibitem[{Shah(1992)}]{shahbook}
Shah, J.
\newblock \emph{Hot Carriers in Semiconductor nanostructures: Physics and
  Applications} (Academic Press, New York, 1992).

\bibitem[{Tr$\ddot{a}$nkle \emph{et~al.}(1987)}]{TranklePRB1987}
Tr$\ddot{a}$nkle, G. \emph{et~al.}
\newblock General relation between band-gap renormalization and carrier density
  in two-dimensional electron-hole plasmas.
\newblock \emph{Phys. Rev. B} \textbf{36}, 6712--6714 (1987).

\bibitem[{Schmitt-Rink \emph{et~al.}(1984)Schmitt-Rink, Ell, Koch, Schmidt \&
  Haug}]{SchmittSSC}
Schmitt-Rink, S., Ell, C., Koch, S.~W., Schmidt, H.~E. \& Haug, H.
\newblock Subband-level renormailization and absorptive optical bistability in
  semiconductor multiple quantum well structures.
\newblock \emph{Solid State Comm.} \textbf{52}, 123--125 (1984).

\bibitem[{Adachi(1985)}]{Adachi}
Adachi, S.
\newblock GaAs, AlAs, and Al$_x$Ga$_{1–x}$As Material parameters for use in
  research and device applications.
\newblock \emph{J. Appl. Phys.} \textbf{58}, R1--R29 (1985).

\end{thebibliography}


\vspace*{0cm}

\vspace*{0cm}

\vspace*{0cm}

\vspace*{0cm}

\vspace*{0cm}

\vspace*{0cm}

\end{document}